\documentclass[aps,longbibliography,showpacs,twocolumn,superscriptaddress]{revtex4-2}
\usepackage{amsmath,amssymb,amsfonts,bm}
\usepackage{graphicx}
\usepackage{epstopdf}
\usepackage{dcolumn}
\usepackage{mathrsfs}
\usepackage{bbold}
\usepackage{dsfont}
\usepackage{float}
\usepackage{ulem}
\usepackage[colorlinks=true,linkcolor=blue,citecolor=blue, urlcolor=blue,bookmarks=false]{hyperref}
\usepackage{tgtermes}

\usepackage{changes}
\colorlet{Changes@Color}{red}

\begin{document}
	
	\title{Floquet Weyl semimetal phases in light-irradiated higher-order topological Dirac semimetals}
	\date{\today }
	\author{Zi-Ming Wang}
	\affiliation{Department of Physics, Hubei University, Wuhan 430062, China}
	\affiliation{Department of Physics and Chongqing Key Laboratory for Strongly Coupled Physics, Chongqing University, Chongqing 400044, China}
	\author{Rui Wang}
	\affiliation{Department of Physics and Chongqing Key Laboratory for Strongly Coupled Physics, Chongqing University, Chongqing 400044, China}
	\affiliation{Center of Quantum Materials and Devices, Chongqing University, Chongqing 400044, China}
	\author{Jin-Hua Sun}
	\email[]{sunjinhua@nbu.edu.cn}
	\affiliation{Department of Physics, Ningbo University, Ningbo 315211, China}
	\author{Ting-Yong Chen}
	\email[]{chenty@sustech.edu.cn}
	\affiliation{Shenzhen Institute for Quantum Science and Engineering, Southern University of Science and Technology, Shenzhen 518055, China}
	\author{Dong-Hui Xu}
	\email[]{donghuixu@cqu.edu.cn}
	\affiliation{Department of Physics and Chongqing Key Laboratory for Strongly Coupled Physics, Chongqing University, Chongqing 400044, China}
	\affiliation{Center of Quantum Materials and Devices, Chongqing University, Chongqing 400044, China}
	
	\begin{abstract}
		
		Floquet engineering, the concept of tailoring a system by a periodic drive, is increasingly exploited to design and manipulate topological phases of matter. In this work, we study periodically driven higher-order topological Dirac semimetals associated with a $k$-dependent quantized quadrupole moment by applying circularly polarized light. The undriven Dirac semimetals feature gapless higher-order hinge Fermi arc states which are the consequence of the higher-order topology of the Dirac nodes. Floquet Weyl semimetal phases with hybrid-order topology, characterized by both a $k$-dependent quantized quadrupole moment and a $k$-dependent Chern number, emerge when illumining circularly polarized light. Such Floquet Weyl semimetals support both hinge Fermi arc states and topological surface Fermi arc states. In addition, Floquet Weyl semimetals with tilted Weyl cones in higher-order topological Dirac semimetals are also discussed. Considering numerous higher-order topological Dirac semimetal materials were recently proposed, our findings can be testable soon.
		
	\end{abstract}
	
	\maketitle
	
	\textit{Introduction.} Understanding Dirac-like fermions has become an imperative in modern
	condensed matter physics. All across the research frontier, ranging from graphene to $d$-wave high-temperature superconductors to topological insulators and beyond, low-energy excitations in various electronic systems can be well-described by the Dirac equation~\cite{Dirac-review}. Of particular interest are Dirac semimetals~(DSMs) as they represent an unusual phase of quantum matter that hosts massless Dirac fermions as quasiparticle excitations near bulk nodal points. Graphene is a well-known two-dimensional (2D) DSM protected by chiral~(sublattice) symmetry~\cite{Graphene-Review}, and stable three-dimensional (3D) DSMs protected by crystalline symmetries had been identified and realized in solid materials as well~\cite{TSM-review}. Due to the lack of bulk-boundary correspondence---the cornerstone of topological phases of matter, the designation of DSMs as a semimetallic topological phase was controversial~\cite{SFAofDSM,wieder2022topological}, whereas DSMs do serve as a parent phase for realizing exotic topological states and topological phenomena. For instance, breaking time-reversal symmetry~(TRS) via magnetism in a DSM can result in the quantum anomalous Hall state~\cite{Haldane1988} or Weyl semimetal~(WSM) states hosting massless chiral fermions and surface Fermi arc states~\cite{WSM-wan}. 
	
	Floquet engineering is a versatile approach that uses time-periodic driving of a quantum system to enable novel out-of-equilibrium many-body quantum states~\cite{RevModPhys.89.011004,RevModPhys.93.041002,weitenberg2021tailoring}. Recent years have witnessed intense efforts toward exploiting Floquet engineering to create topological phases in quantum materials~\cite{pssr.201206451,annurev-conmatphys-031218-013423,harper-annurev-conmatphys,rudner2020band,bao2022light,PhysRevB.79.081406,PhysRevLett.105.017401,PhysRevB.82.235114,PhysRevLett.107.216601,PhysRevB.84.235108,PhysRevB.88.245422,PhysRevLett.113.266801,PhysRevB.90.115423,lindner2011floquet,PhysRevA.82.033429,PhysRevB.87.235131,PhysRevLett.106.220402,PhysRevX.3.031005,PhysRevLett.110.026603,PhysRevLett.110.200403,PhysRevLett.112.156801,Wang_2014,PhysRevB.94.041409,PhysRevLett.117.087402,PhysRevB.94.155206,PhysRevLett.116.156803,PhysRevB.91.205445,PhysRevB.94.081103,PhysRevB.94.121106,PhysRevB.96.041206,PhysRevB.93.205435,hubener2017creating,PhysRevB.97.155152,PhysRevB.97.195139,PhysRevB.100.165302,PhysRevB.102.201105,Ghosh2022systematic,nag2021anomalous,Nag2019dynamical,XLDU2022}. It is important to stress that circularly polarized light~(CPL) naturally breaks TRS, which provides an easy tuning knob to induce dynamical topological phases such as Floquet Chern insulators and Floquet WSMs in DSMs~\cite{PhysRevB.79.081406,PhysRevLett.105.017401,PhysRevLett.107.216601,PhysRevB.84.235108,PhysRevB.88.245422,PhysRevLett.113.266801,PhysRevB.90.115423,PhysRevLett.116.156803,PhysRevB.94.121106,hubener2017creating,PhysRevLett.128.066602,Wang453,mciver2020light}. 
	Following the discovery of the concept of higher-order topology that characterizes boundary states with dimensions two or more lower than that of the bulk system that accommodates them ~\cite{Zhang2013PRL,Benalcazar2017Science,Langbehn2017PRL,Song2017PRL,Schindler2018SA,PhysRevLett.124.036803,xie2021higher}, there has been a surge of interest in tailoring higher-order topological phases by using Floquet engineering as well~\cite{PhysRevB.99.045441,PhysRevB.100.085138,PhysRevResearch.1.032045,PhysRevB.100.115403,PhysRevResearch.1.032045,PhysRevLett.123.016806,PhysRevLett.124.057001,PhysRevB.101.235403,PhysRevResearch.2.013124,PhysRevResearch.2.033495,PhysRevLett.124.216601,PhysRevB.103.L041402,PhysRevB.103.L121115,PhysRevB.103.115308,PhysRevB.104.L020302,PhysRevResearch.3.L032026,PhysRevB.103.184510,PhysRevB.104.205117,Ghosh2022hinge-mode}. Meanwhile, numerous higher-order topological DSMs, which obey the topological bulk-hinge correspondence and thus display universal topological hinge Fermi arc states, have been proposed~\cite{Mao2018PRB,wieder2020strong,yang2021classification,Tyner2021quantized,Nie2022HODSM,zeng2022topological}. The signature of hinge Fermi arc states was recently observed in supercurrent oscillation experiments on prototypical DSM material Cd$_3$As$_2$~\cite{HOTDSM-EXP,WANG2022788}. Cd$_3$As$_2$ provides a promising parent material for the realization of the higher-order WSM~\cite{Wang2020higher-order,Ghorashi2020higher-order,Roy2019antiunitary} that supports hinge Fermi arc states in addition to the usual surface Fermi arc states by using Floquet engineering.
	
	In this work, we explore tunable higher-order WSMs in time-symmetric and $PT$-symmetric higher-order topological DSMs under off-resonant CPL illumination. Without driving, both types of higher-order topological DSMs have two bulk Dirac nodes locating at the $k_z$-axis and support gapless hinge Fermi arc states. Meanwhile, the time-symmetric one has additional closed surface Fermi rings. CPL drives each Dirac node to split into a pair of Weyl nodes by symmetry breaking, resulting in Floquet higher-order WSMs accommodating rich topological boundary states. The coexistence of surface Fermi arc and hinge Fermi arc states signals a hybrid-order topology which can be captured by $k$-dependent quantized quadrupole moment and Chern number. In addition, the surface Fermi rings in the time-symmetric DSM are inherited in the Floquet WSM. Moreover, we can achieve a type-II higher-order WSM with overtilted Weyl cones by adjusting the incident direction of CPL. Our proposal can be realized in DSM materials like Cd$_3$As$_2$ with current ultrafast experimental techniques.
	
	\textit{DSM model and the Floquet theory.} Undriven higher-order topological DSMs are constructed based on a generic band inversion DSM model on the cubic lattice. In reciprocal space, the Hamiltonian matrix is
	\begin{align}
		H_\text{}(\bm{k}) \nonumber &=\epsilon_{0}(\mathbf{k})+\lambda\sin{k_x} \Gamma_{1}+\lambda\sin{k_y} \Gamma_{2} \\ 
		&+M(\mathbf{k}) \Gamma_{3}+G(\mathbf{k})\Gamma_{4} ,
		\label{eq1}
	\end{align}
	where the $\Gamma$ matrices are $\Gamma_{1}=s_{3}\sigma_{1}$, $\Gamma_{2}=s_{0}\sigma_{2}$, $\Gamma_{3}=s_{0}\sigma_{3}$, $\Gamma_{4}=s_{1}\sigma_{1}$, and $\Gamma_{5}=s_{2}\sigma_{1}$, with $s_{j=1,2,3}$ and $\sigma_{j=1,2,3}$ the Pauli matrices labeling the spin and orbital degrees of freedom, respectively, and $s_{0}$, $\sigma_{0}$ are $2 \times 2$ identity matrices. $\epsilon_{0}(\mathbf{k})=t_1(\cos{k_z}-\cos K_z^0)+t_2(\cos{k_x}+\cos{k_y}-2)$, and $M(\mathbf{k})=t(\cos{k_x}+\cos{k_y}-2)+t_z(\cos{k_z}-\cos K_z^0)$. $t_{1,2}$, $t$, $t_z$, and $\lambda$ are the amplitudes of hoppings. The DSM has two Dirac points locating at $\mathbf{k}^0=(0,0,\pm K_z^0)$. $G(\mathbf{k})$ is the coefficient of $\Gamma_4$ term that gives birth to higher-order topology in the present DSMs. Without the $\Gamma_4$ term, this model describes an ordinary band-inversion DSM supporting helical surface Fermi arc states which are not topologically stable~\cite{SFAofDSM}. Equation (\ref{eq1}) can describe the recently identified higher-order DSM materials, including but not limited to Cd$_3$As$_2$ and KMgBi~\cite{wieder2020strong}.
	
Treating $k_z$ as a parameter, then $H_{}({\mathbf{k}})$ reduces to a 2D Hamiltonian $H_{k_z}(k_x,k_y)$. The reduced Hamiltonian possesses higher-order topology which can be well characterized by the quantized quadrupole moment. Furthermore, we can use a $k_z$-dependent quantized quadrupole moment $\mathcal{Q}_{xy}(k_z)$ to capture the higher-order topology of the DSMs. $\mathcal{Q}_{xy}$ in real space~\cite{Wheeler2019manybody,Kang2019manybody} is
	\begin{equation}
		\mathcal{Q}_{xy}=\frac{1}{2\pi}\renewcommand{\Im}{\operatorname{Im}}  \Im[\log{\langle\hat{U}_{xy}\rangle}], \quad
		\hat{U}_{xy}=e^{i2\pi\sum_{\bm{r}_i}\hat{q}_{xy}(\bm{r}_i)},
		\label{eq3}
	\end{equation}
	where $\hat{q}_{xy}(\bm{r}_i)=\frac{xy}{L_{x} L_{y}}\hat{n}(\bm{r}_i)$ is the quadrupole moment density per unit cell at site $\bm{r}_i$, and $L_{x}$ and $L_{y}$ are the length of the system in the $x$ and $y$ directions, respectively.
	
	\begin{figure}[h!tpb]
	\includegraphics[width=8cm]{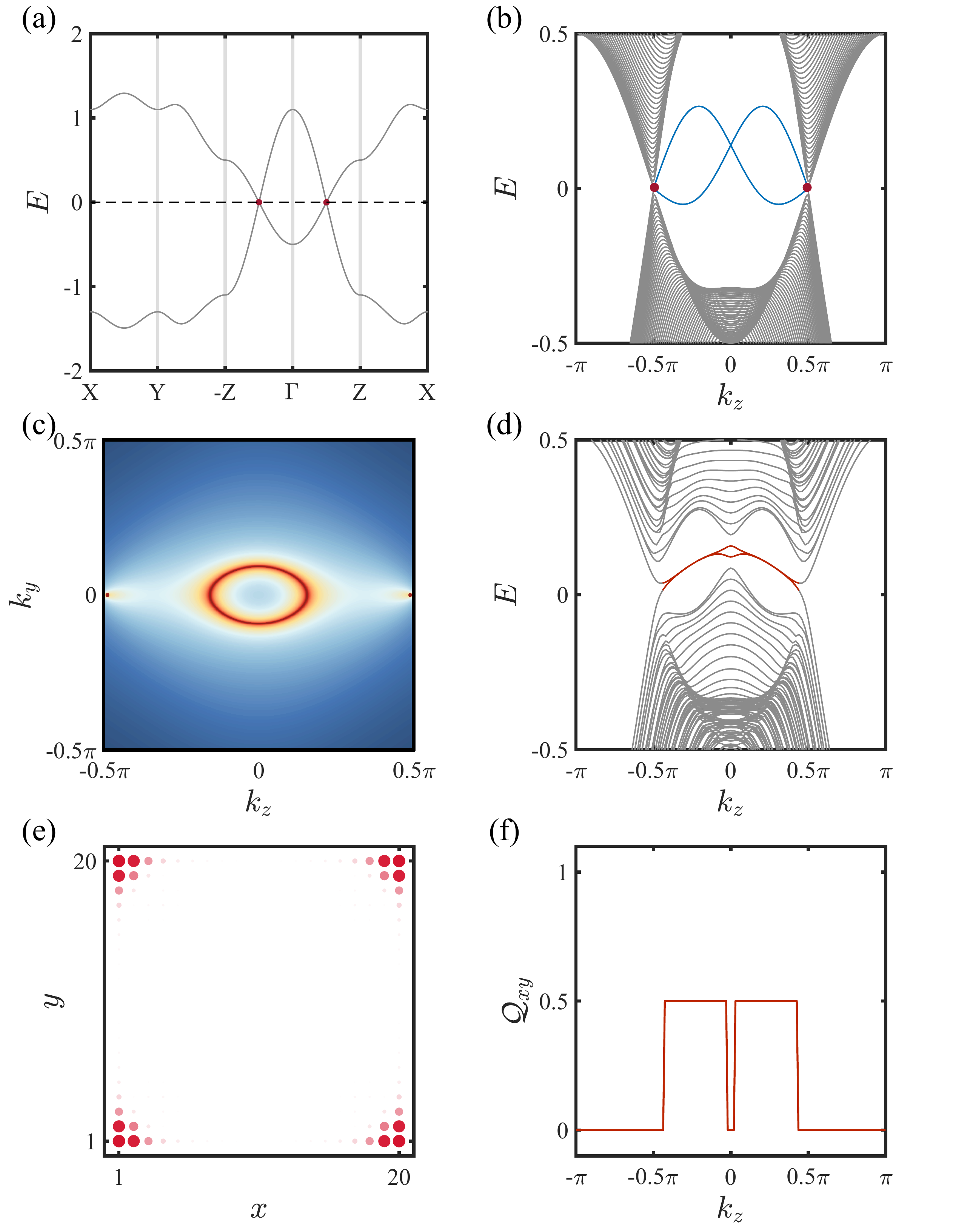}
	\caption{Electronic structures and bulk topology of the time-symmetric DSM. (a) Bulk band structure along high-symmetry points in the 3D Brillouin zone. The red dots mark the bulk Dirac points. (b) The surface band dispersion versus $k_z$ direction for $k_y=0$. The open boundary condition (OBC) is imposed along the $x$-direction. The solid blue lines show the gapless surface Dirac states. (c) The surface spectral function on the $k_y$-$k_z$ plane with the semi-infinite boundary along the $x$-direction when $E=0$. The red dots at $\pm\pi/2$ are the projection of Dirac points, and the red circle marks the surface Fermi rings from the surface Dirac states shown in (b). (d) The energy spectrum versus $k_z$ with the OBC along both the $x$ and $y$ directions. The solid red lines represent the topologically protected hinge Fermi arc states. The Dirac points at $k_z=\pm \pi/2$ are gapped due to the size effect. (e) The local density of states (LDOS) of the hinge Fermi arc states at $k_{z}=0.1\pi$. (f) The $k_{z}$-dependent quantized quadrupole moment $\mathcal{Q}_{xy}$. The two quantized plateaus of $Q_{xy}$ correspond to two segments of degenerate hinge Fermi arc states. The parameters are chosen as $t_1=0.3$, $t_2=0.2$, $\lambda=0.5$, $t=1$, $t_z=0.8$, $K_{z}^{0}=\pi/2$, and $g=-0.4$.}
	\label{Figure1}
    \end{figure}			
	
	CPL is described by a time-periodic gauge field $\mathbf{A}(\tau)=\mathbf{A}(\tau+T)$ with period $T=\frac{2\pi}{\omega}$ and the frequency of light $\omega$. In specific, the gauge field $\mathbf{A}=A(0,\eta\sin\omega\tau,\cos\omega\tau), A(\cos\omega\tau, 0, \eta\sin\omega\tau)$, and $A(\eta\sin \omega \tau, \cos\omega\tau,0)$, where $\eta=\pm1$ labeling the handedness, describe CPL propagating along the $x$, $y$ and $z$ directions, respectively. In the main text, we mainly focus on the case of CPL propagating along the $z$-direction, the case of CPL propagating along the $x$-direction is also studied in the Supplemental Material~\cite{Supplement}. Electrons moving on a lattice couple to the electromagnetic gauge field via the Peierls substitution: $\tilde{t}\rightarrow \tilde{t}\exp[-i\int_{\mathbf{r}_{j}}^{\mathbf{r}_{k}}\mathbf{A}(\tau)\cdot d\mathbf{r}]$, where $\mathbf{r}_j$ is the coordinate of lattice site $j$. Thereby, in the presence of CPL, the DSM Hamiltonian becomes periodic in time $H(\tau)=H(\tau+T)$. In the following, we use the natural units $e=\hbar=c=1$. Thanks to Floquet's theorem, we can transfer the time-dependent Hamiltonian problem to a time-independent one~\cite{PhysRev.138.B979,PhysRevA.7.2203}. In specific, the time-dependent Schr\"{o}dinger equation has a set of solution $|\Psi(\tau) \rangle =e^{-i\epsilon \tau}|\Phi(\tau)\rangle$, where $\epsilon$ denotes the Floquet quasienergy, and $|\Phi(\tau)\rangle=|\Phi(\tau+T)\rangle$ is dubbed the Floquet state. Expanding the Floquet state in a Fourier series  $|\Phi(\tau)\rangle=\sum_n e^{-in\omega \tau}|\Phi^n\rangle$, we arrive at an infinite dimensional eigenvalue equation in the extended Hilbert (or Sambe) space
	\begin{equation}
		\sum_m(H_{n-m}-m\omega \delta_{mn})|\Phi_\alpha^m\rangle=\epsilon_\alpha|\Phi_\alpha^m\rangle,
	\end{equation}
	where $H_{n-m}=\frac{1}{T}\int_{0}^{T} d \tau e^{i(n-m)\omega\tau} H(\tau)$. Throughout, we focus on the case in the high frequency limit, where the resonant interband transitions are very unlikely. This case yields an effective static Floquet Hamiltonian~\cite{bukov2015universal,Eckardt_2015}
	\begin{equation}
		H_{\text{eff}}=H_{0}+\sum_{l\neq 0}\frac{\left[ H_{-l},H_{l}\right] }{l\omega }+\mathcal O(\omega ^{-2}).
		\label{eq.4}
	\end{equation}
	In our calculations, the maximum value of $l$ is determined by checking whether the results converge. 
		
	\textit{Light-irradiated time-symmetric higher-order topological DSM.} First, we specify $G(\mathbf{k})=g(\cos{k_x}-\cos{k_y})\sin{k_z}$, which breaks the fourfold rotation symmetry $\mathcal{C}_{4z}$ but preserves TRS $\mathcal{T}$ and inversion symmetry $\mathcal{P}$, where $\mathcal{P}=\Gamma_{3}$ and $\mathcal{T}=is_2\mathcal{K}$ with $\mathcal{K}$ the complex conjugation. With the present form of $G(\mathbf{k})$, the Dirac points shown in Fig.~\ref{Figure1}(a) stay at the same position as the case without $G(\mathbf{k})$. Yet, the surface states behave quite differently. A closed Fermi ring instead of helical Fermi arc states [Fig.~\ref{Figure1}(c)], emerges in the surface Brillouin zone~\cite{SFAofDSM} when the Fermi energy cut through the surface Dirac cone [see the band crossing around $k_z=0$ marked in blue in Fig.~\ref{Figure1}(b)]. More importantly, the DSM is endowed with the higher-order topology characterized by $\mathcal{Q}_{xy}(k_z)$ shown in Fig.~\ref{Figure1} (f). Note that, $G(k_x,k_y,k_z=0)=0$ due to explicit dependence of $\sin k_z$, which results in a vanishing $\mathcal{Q}_{xy}$ within a small window around $k_z=0$ for a finite-size system. Quantized $\mathcal{Q}_{xy}$ leads to topological hinge Fermi arcs terminated by the projection of the Dirac points, as shown in Figs.~\ref{Figure1}(d) and \ref{Figure1}(e). Recently, the evidence of hinge Fermi arc states was experimentally reported in DSM Cd$_3$As$_2$~\cite{HOTDSM-EXP,WANG2022788}, which might be attributed to the existence of this type of $G(\mathbf{k})$. 
	
	\begin{figure}[h!tpb]
	\includegraphics[width=8cm]{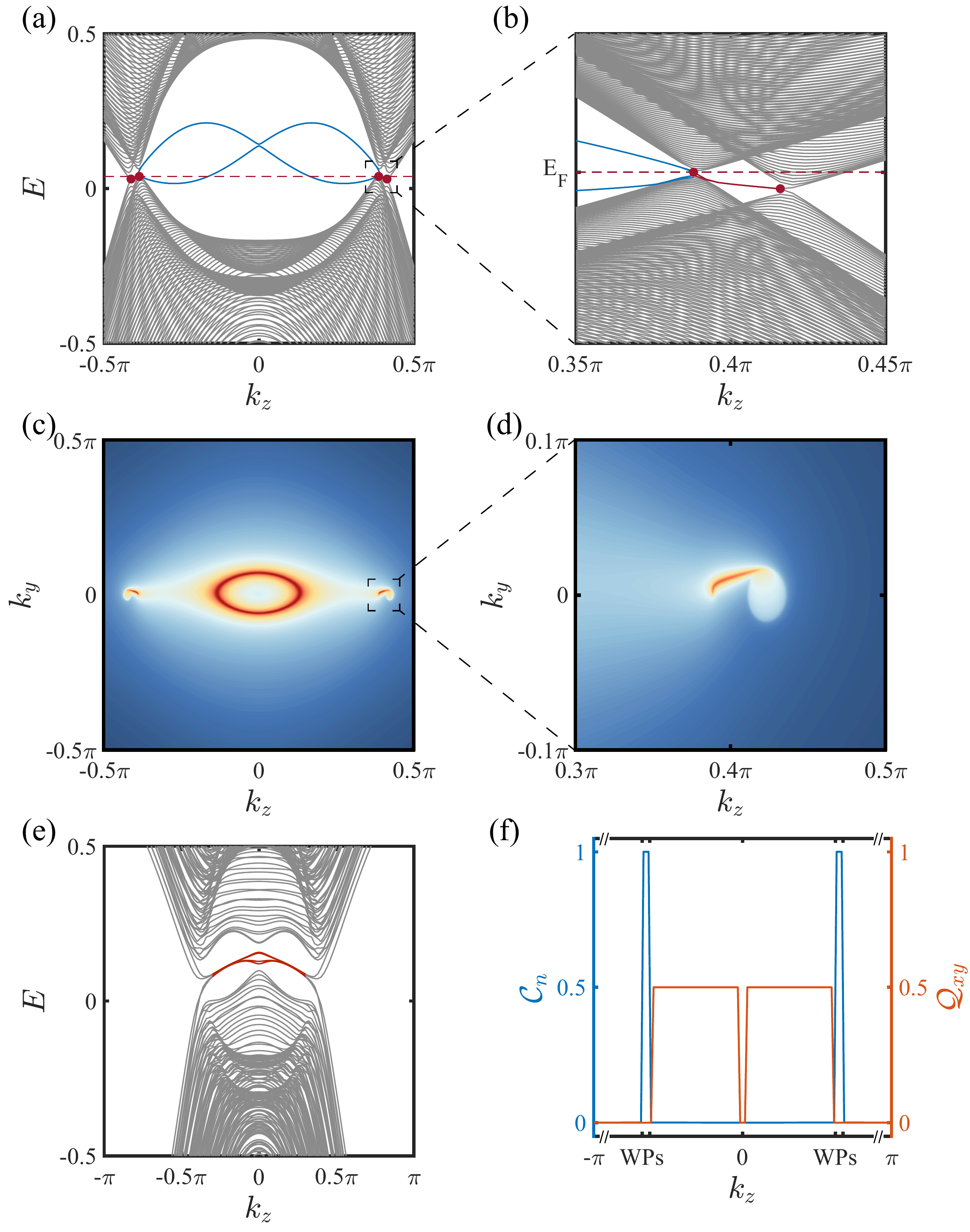}
	\caption{The electronic structure and bulk topology of the Floquet WSM in the light-irradiated time-symmetric DSM. (a) The surface band structure versus $k_z$ for $k_y=0$. (b) The zoom-in view of the area around the pair of Weyl points within the black dashed square in (a). The red dots show the Weyl points, and the red solid lines represent the surface Fermi arc states. (c) The surface spectral function on the $k_y$-$k_z$ plane with the semi-infinite boundary along the $x$ direction at $E_\text{F}$ marked in red dashed line in (b). The red solid lines around the original Dirac points mark the surface Fermi arcs, which connect the same pair of Weyl points. The red closed circle shows the surface Fermi rings. (d) The zoom-in view of the surface Fermi arc within the black dashed square in (c). The shaded area connecting the right end of the Fermi arc marks the electron pocket. (e) The energy spectrum as a function of $k_{z}$ for the OBC along the $x$ and $y$ directions. The hinge states are marked by the solid red lines. (f) $\mathcal{Q}_{xy}$ and the Chern number $\mathcal{C}_n$ marked in orange and blue, respectively, as functions of $k_z$. The parameters are the same as those in Fig.~\ref{Figure1} except $A=0.7$, and $\omega=3$.}
	\label{Figure2}
\end{figure}

	In the presence of CPL propagating along the $z$ direction, we obtain an effective Floquet Hamiltonian according to Eq.~(\ref{eq.4})
	\begin{align}
		&H_\text{eff}^\text{I}(\bm{k}) \nonumber =[t_1(\cos{k_z}-\cos{K_z^0})+t_2\mathcal{J}_{0}(A)(\cos{k_x}+\cos{k_y}) \\ \nonumber
		&-2 t_2] + \lambda\mathcal{J}_{0}(A)(\sin{k_x}\Gamma_{1}+\sin{k_y}\Gamma_{2})+\mathcal{J}_{0}(A)G(k)\Gamma_{4}\\ \nonumber 
		&\!+\![t_z(\cos{k_z} \!-\!\cos{K_z^0})\!+\!t\mathcal{J}_{0}(A)(\cos{k_x} \!+\!\cos{k_y})-2 t]\Gamma_{3}  \\ \nonumber
		& +\sum\limits_{l>0,l\in odd}\frac{4\eta\mathcal{J}_{l}^{2}(A)}{l\hbar\omega}\{\lambda^2\cos{k_x}\cos{k_y}\Gamma_{12} \\ \nonumber
		& -\lambda t [\cos{k_x}\sin{k_y}\Gamma_{13}-\sin{k_x}\cos{k_y}\Gamma_{23}] \\ \nonumber
		& +\lambda g \sin{k_z}[\cos{k_x}\sin{k_y}\Gamma_{14}-\sin{k_x}\cos{k_y}\Gamma_{24}] \\ 
		& +2t g\sin{k_x}\sin{k_y}\sin{k_z}\Gamma_{34}\},
		\label{eq.5}
	\end{align}
	where $\mathcal{J}_{l}(A)$ is the Bessel function of $l$th order and of the first kind, $\Gamma_{j k}=[\Gamma_{j},\Gamma_{k}]/2i$. Equation~(\ref{eq.5}) indicates that CPL not only renormalizes electron hopping perpendicular to the propagation direction but also induces TRS breaking hopping terms.
	Notably, each Dirac point is separated into a pair of Weyl points, resulting in a Floquet WSM with two pairs of Weyl nodes.
	The Floquet WSM hosts surface Fermi arc states around the original Dirac points, which connect the projection of each single pair of Weyl nodes as depicted in Figs.~\ref{Figure2}(a)-\ref{Figure2}(d). Note that, electron pockets are formed around the projection of two of the four Weyl points because each pair of Weyl nodes are separated in energy [see the zoom-in plots in Figs.~\ref{Figure2}(b) and \ref{Figure2}(d)]. As depicted in Figs.~\ref{Figure2}(a) and \ref{Figure2}(c), the WSM also supports the closed surface Fermi ring generated by surface Dirac states as does the undriven DSM. Coexistence of Fermi arcs and Fermi rings on the surface of time-reversal invariant WSM had been revealed in Ref.~\cite{Alexanderprl2017}. In our case, TRS is broken in the Floquet WSM, however, the coexisting Fermi arcs and Fermi ring are still present. Moreover, the Floquet WSM also features the hinge Fermi arc states terminated by the projection of two adjacent Weyl nodes from two different pairs~[Fig.~\ref{Figure2}(e)]. To capture the topology of the Floquet WSM, in Fig.~\ref{Figure2}(f), we show the quantized quadrupole moment and the Chern number as functions of $k_z$. We can see that the quadrupole moment takes a quantized value of $\mathcal{Q}_{xy}=1/2$ in between the two middle Weyl points, and the Chern number displays $\mathcal{C}_n=1$ between the same pairs of Weyl nodes. 
	
    The Dirac points in Cd$_3$As$_2$ come from the band crossing induced by band inversion. The characteristic low-energy scale describing the band inversion is about 50 meV \cite{Arribi2020topological,Wang2013threedimensional}. We notice the pump photon energy in a recent experiment on Floquet band engineering can reach 440 meV~\cite{zhou2023pseudospin}. Therefore, the implementation of CPL at high frequency is feasible in such a Dirac semimetal material. 
	
   \begin{figure}[h!tpb]
     \includegraphics[width=8cm]{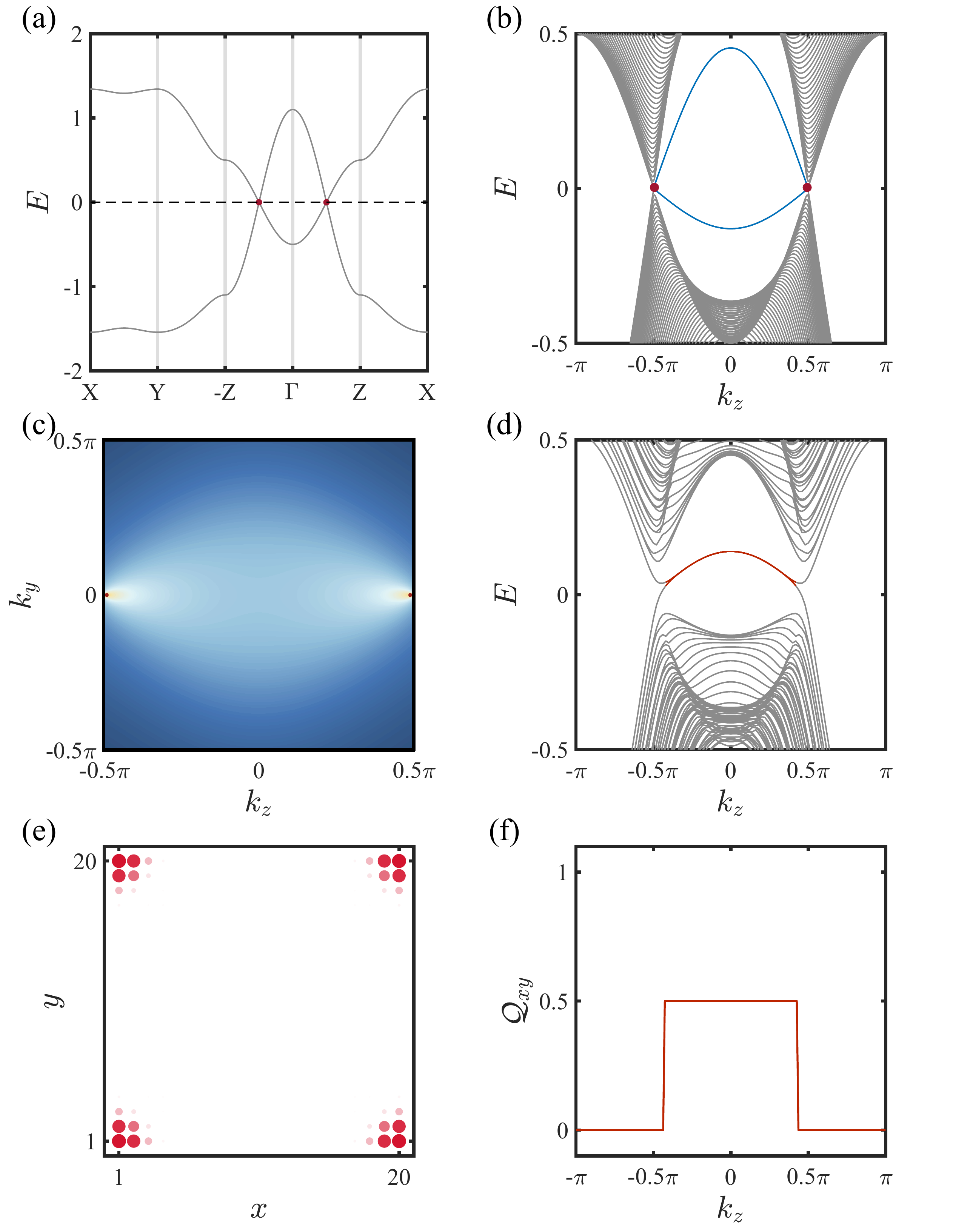}
     \caption{The energy spectrum and bulk topology of the $PT$-symmetric DSM. (a) Band structure along high-symmetry points in the 3D Brillouin zone. (b) The surface band structure versus $k_z$ direction for $k_y=0$. The blue solid lines show the separable surface state between two Dirac points marked in red dots. (c) The surface spectral function on the $k_y$-$k_z$ plane with the semi-infinite boundary along the $x$-axis when $E=0$. The red dots are the projection of two Dirac points. (d) The energy spectrum versus $k_{z}$ for the OBC along both the $x$ and $y$ directions. The solid
     red lines represent the topologically protected hinge Fermi arc states. (e) The LDOS of the hinge Fermi arc states at $k_{z}=0.1\pi$. (f) The $k_{z}$-dependent quantized quadrupole moment $Q_{xy}$. $Q_{xy}$ shows a quantized plateau where the hinge Fermi arc marked in (d) exist. The parameters are chosen as $t_1=0.3$, $t_2=0.2$, $\lambda=0.5$, $t=1$, $t_z=0.8$, $K_{z}^{0}=\pi/2$, and $g=-0.4$.}
     \label{Figure3}
     \end{figure}
     
	\begin{figure}[h!tpb]
	\includegraphics[width=8cm]{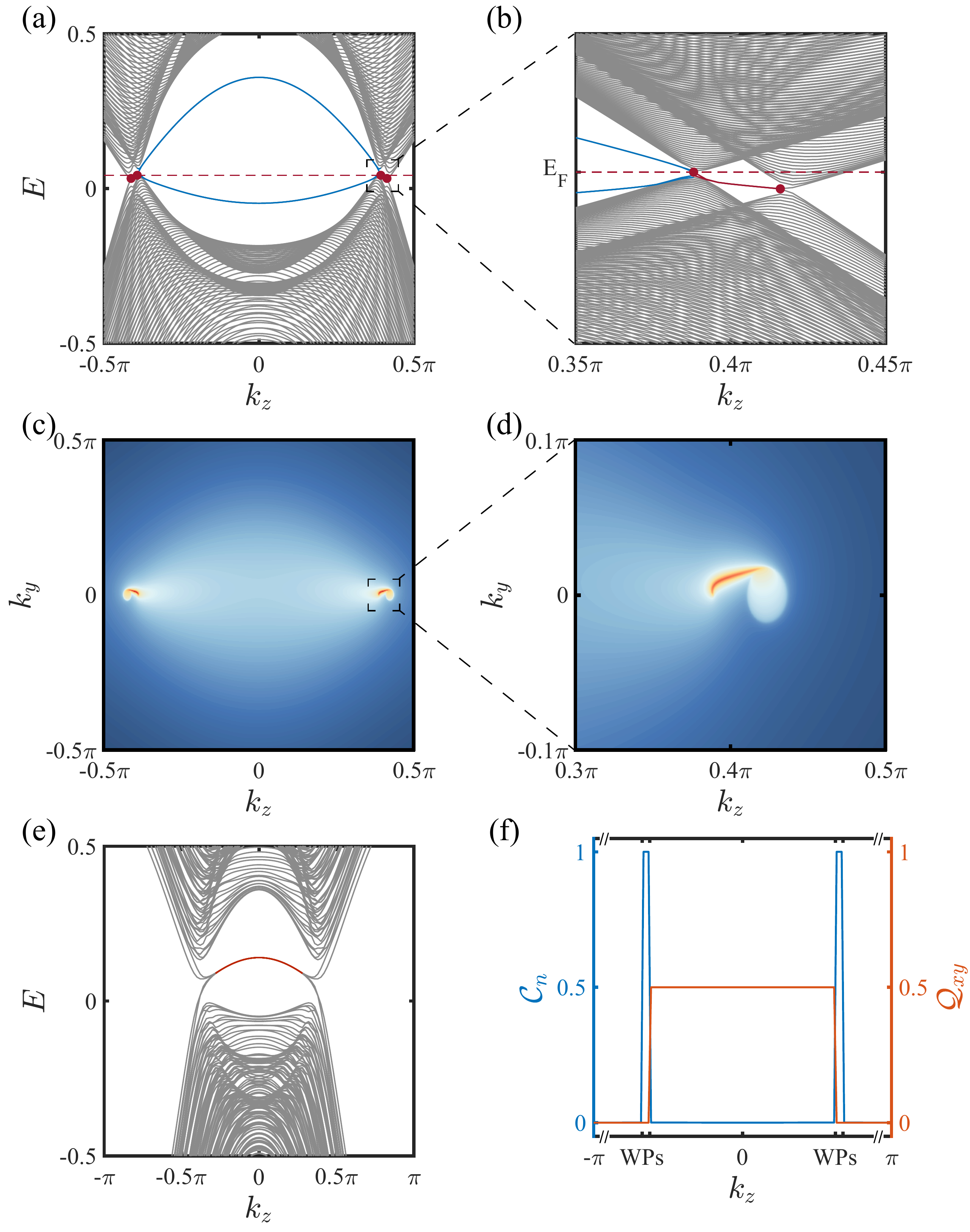}
	\caption{The electronic structure and bulk topology of the Floquet WSM in the light-irradiated $PT$-symmetric DSM. (a) The surface band structure along $k_z$ direction when $k_y=0$. The blue solid lines show the separable surface state between the two middle of Weyl points marked in red dots. (b) The zoom-in view of the area around the pair of Weyl points within the black dashed square in (a). The red solid lines represent the surface Fermi arc states. (c) The surface spectral function on the $k_y$-$k_z$ plane with the semi-infinite boundary along the $x$ direction at $E_\text{F}$ marked in red dashed line in (b). The red lines around the original Dirac points show the surface Fermi arc states, which connect the same pair of Weyl points. The zoom-in view of the surface Fermi arc within the black dashed square in (c). The shaded area attached the right end of the Fermi arc marks the electron pocket. (e) The energy spectrum as a function of $k_{z}$ for the OBC along the $x$ and $y$ directions. (f) $\mathcal{Q}_{xy}$ and $\mathcal{C}_n$ marked in orange and blue, respectively, as functions of $k_z$. The parameters are the same as those in Fig.~\ref{Figure2}.}
		\label{Figure4}
	\end{figure}
	\textit{Light-irradiated $PT$-symmetric higher-order topological DSM.} We consider $G(\mathbf{k})$ is independent of $k_z$, i.e., $G(\mathbf{k})=g(\cos k_x-\cos k_y)$, it also generates higher-order topology in the DSM. The previously defined TRS is broken, however, the system still has two four-fold degenerate Dirac points [Fig.~\ref{Figure3}(a)], which is called $PT$-symmetric DSM as the combination of TRS and inversion symmetry $\mathcal{P}\otimes\mathcal{T}=is_2\sigma_2\mathcal{K}$ is preserved. It is still of interest to ask whether CPL induces intriguing Floquet topological states as it does in the time-symmetric DSM. Before going into the Floquet states, it's better to take a look at the electronic structure of the DSM. From Fig.~\ref{Figure3}, we can see that, in contrast to the time-symmetric type DSM, the surface states in $PT$-symmetric DSM are separated, and no surface Dirac cones exist [Figs.~\ref{Figure3}(b) and \ref{Figure3}(c)]. Still, it supports gapless hinge Fermi arc states terminated on the projection of two Dirac points as shown in Figs.~\ref{Figure3}(d) and \ref{Figure3}(e). Moreover, in Fig.~\ref{Figure3}(f), we calculate $\mathcal{Q}_{xy}(k_z)$ to capture the higher-order topology of this type of DSM. Such a type of higher-order topological DSM can be realized in antiferromagnetic systems as the present $\Gamma_{4}$ term may represent an orbital-dependent spin density wave.
	
	\begin{figure}[h!tpb]
	\includegraphics[width=8cm]{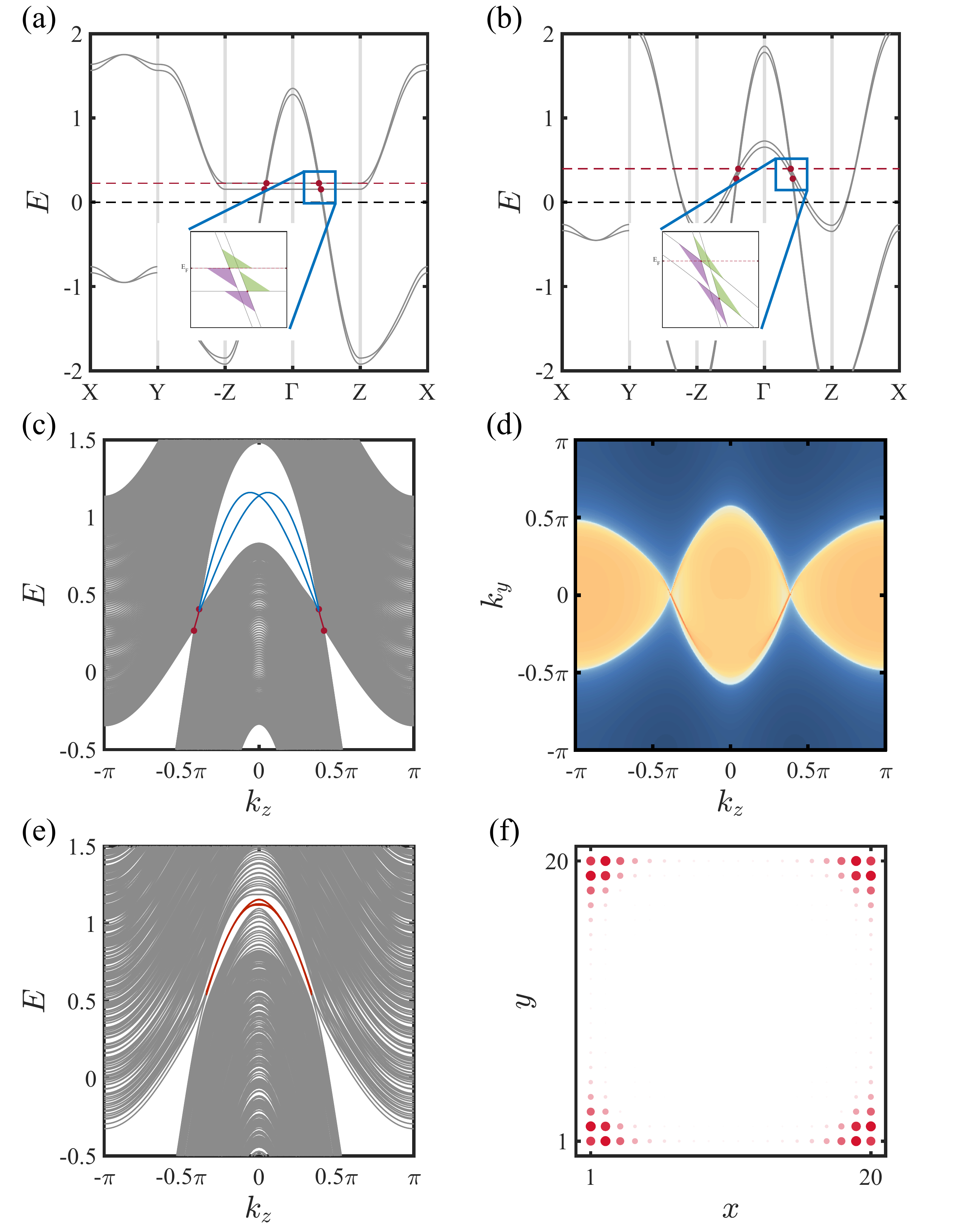}
	\caption{The energy spectrum of tilted Floquet WSMs. (a) Bulk band structure of the critical type of WSM along the high-symmetry points in the 3D Brillouin zone. The four red dots mark the Weyl nodes. The inset is a zoom of a pair of Weyl nodes. The green and purple triangles in the inset represent the conduction band and valence band, respectively. (b) The bulk band of the type-II WSM with $t_1=1.3$ and $| t_1 / t_z |>1$. (c) The surface band structure along $k_z$ direction when $k_y=0$. The solid blue lines show the gapless surface Dirac states, and red lines are the surface Fermi arc states between pairs of Weyl nodes. (d) The surface spectral function on the $k_y$-$k_z$ plane with the OBC along the $x$-axis when $E=0.29$. The electron and hole pockets are marked by yellow shaded regions. The red curves are the surface Fermi arcs. (e) The energy spectrum versus $k_{z}$ for the OBC along the $x$ and $y$ directions. The red solid lines represent the hinge Fermi arc state. (f) The LDOS of the hinge Fermi arc state at $k_{z}=0.1\pi$. The parameters are chosen to be $t_2=0.2$, $\lambda=0.5$, $t=1$, $t_z=0.8$, $K_{z}^{0}=\pi/2$, $g=-0.4$, $A=0.7$, and $\omega=3$.}
	\label{Figure5}
	\end{figure}
	Under the high frequency CPL along the $z$ direction, we also obtain an effective Floquet Hamiltonian for the $PT$-symmetric higher-order topological DSM, which can be considered as a special case of Eq.~(\ref{eq.5}) where $\sin k_z=1$. Similarly, CPL renormalizes the original hoppings and also induces the next-nearest neighbor hoppings on the $x$-$y$ plane which break $\mathcal{P}\otimes\mathcal{T}$. Accordingly, each Dirac point splits into a pair of Weyl points along the $k_z$ direction in reciprocal space. Interestingly, the two middle Weyl points from two different pairs are connected by two separable surface state bands [see the solid blue lines in Fig.~\ref{Figure4}(a)], while the two Weyl points within the same pair evolved from the same Dirac point are connected by surface Fermi arc states, as displayed in Figs.~\ref{Figure4}(b), \ref{Figure4}(c) and \ref{Figure4}(d). Just like in the light-irradiated time-symmetric DSM, the surface Fermi arc states are also terminated by the projection of a Weyl node and an electron pocket. The emergence of the Fermi arc states is attributed to a $k$-dependent $\mathcal{C}_n$ plotted in Fig.~\ref{Figure4}(f). In addition, as shown in Fig.~\ref{Figure4}(e), gapless hinge Fermi arc states exist between the two middle Weyl points, whose higher-order topology can also be characterized by $\mathcal{Q}_{xy}(k_z)$ demonstrated in Fig.~\ref{Figure4}(f).

	\textit{Tilting effect on the Floquet WSM.} KMgBi was identified as a critically tilted DSM~\cite{le2017three-dimensional} with  higher-order topology~\cite{wieder2020strong} that is believed to be caused by $G(\mathbf{k})=g(\cos{k_x}-\cos{k_y})\sin{k_z}$. Strain or chemical doping can change the tilt of Dirac cones and drive the DSM into a phase with type-II~(i.e., overtilted)~\cite{soluyanov2015type,xu2015structed} Dirac cones. The transition between type-I and type-II Dirac cones is determined by the ratio of $| t_1 / t_z |$ in Eq.~(\ref{eq1}). For over-titled Dirac cones with $| t_1 / t_z |>1$, shining CPL can also create Floquet WSM states. Figure \ref{Figure5} shows the Floquet WSM in the CPL-illuminated time-symmetric DSM with tilted Dirac cones. When $| t_1 / t_z |=1$, the DSM is at the critical point between the type-I and type-II DSMs. Accordingly, a critical type of Floquet WSM is obtained as shown in Fig.~\ref{Figure5}(a). When $| t_1 / t_z |>1$, a type-II Floquet WSM appears by applying CPL. In this case, electron and hole pockets meet at the Weyl points [Fig.~\ref{Figure5}(b)]. From Figs.~\ref{Figure5}(c)-\ref{Figure5}(f), we can see that the type-II WSM also hosts surface Dirac states, surface Fermi arc states as well as hinge Fermi arc states, however, these boundary states are deeply buried in the bulk states.
	
    At last, we would like to briefly discuss how the Floquet WSM responds to change in propagation orientation of CPL. Applied CPL propagating along the $x$-direction can also induce Floquet WSM states in higher-order topological DSMs. In the Supplemental Material~\cite{Supplement}, we show that starting with a time-symmetric higher-order topological DSM with type-I Dirac cone, the CPL along the $x$-direction can induce a Floquet WSM with over-tilted type-II Weyl cones. It implies that the type of Floquet WSMs can be tuned by adjusting the propagation direction of incident light.
	
	\textit{Conclusions.} In this work, we have investigated the CPL-induced Floquet WSM states in both time-symmetric and $PT$-symmetric higher-order topological DSMs. The emergent Floquet WSM states exhibit higher-order topological hinge Fermi arc states characterized by a $k$-dependent quadrupole moment. Moreover, the Floquet WSMs show gapless Fermi arc states whose topology is characterized by a $k$-dependent Chern number. Our work not only reveals exotic hybrid-order topological WSM states but also suggests an approach to realizing such WSMs in solids rather than acoustic systems~\cite{luo2021observation,wei2021higher}. 
	
	\textit{Acknowledgments.} The authors acknowledge the support by the NSFC (under Grant Nos. 12074108, 12222402, and 12147102), the Natural Science Foundation of Chongqing (Grant No. CSTB2022NSCQ-MSX0568) as well as Shenzhen Institute for Quantum Science and Engineering (under Grant No. SIQSE202101).

	\bibliography{bibfile}

	

	\pagebreak
	\widetext
	\clearpage
	\setcounter{equation}{0}
	\setcounter{figure}{0}
	\setcounter{table}{0}
	\makeatletter
	\renewcommand{\figurename}{FIG.}
	\renewcommand{\theequation}{S\arabic{equation}}
	\renewcommand{\thetable}{S\arabic{table}}
	\renewcommand{\thefigure}{S\arabic{figure}}

	\begin{center}
		\textbf{Supplemental Material to: ``Floquet Weyl semimetal phases in light-irradiated higher-order topological Dirac semimetals''}
	\end{center}

		In this supplemental material, we present the Floquet Weyl semimetal (WSM) phases induced by the circularly polarized light (CPL) propagating along the $x$-direction and its tilting effect. 
		\section{Floquet WSM induced by CPL propagating along the $x$-direction and its tilting effect}
\begin{figure}[h!tpb]
			\includegraphics[width=11cm]{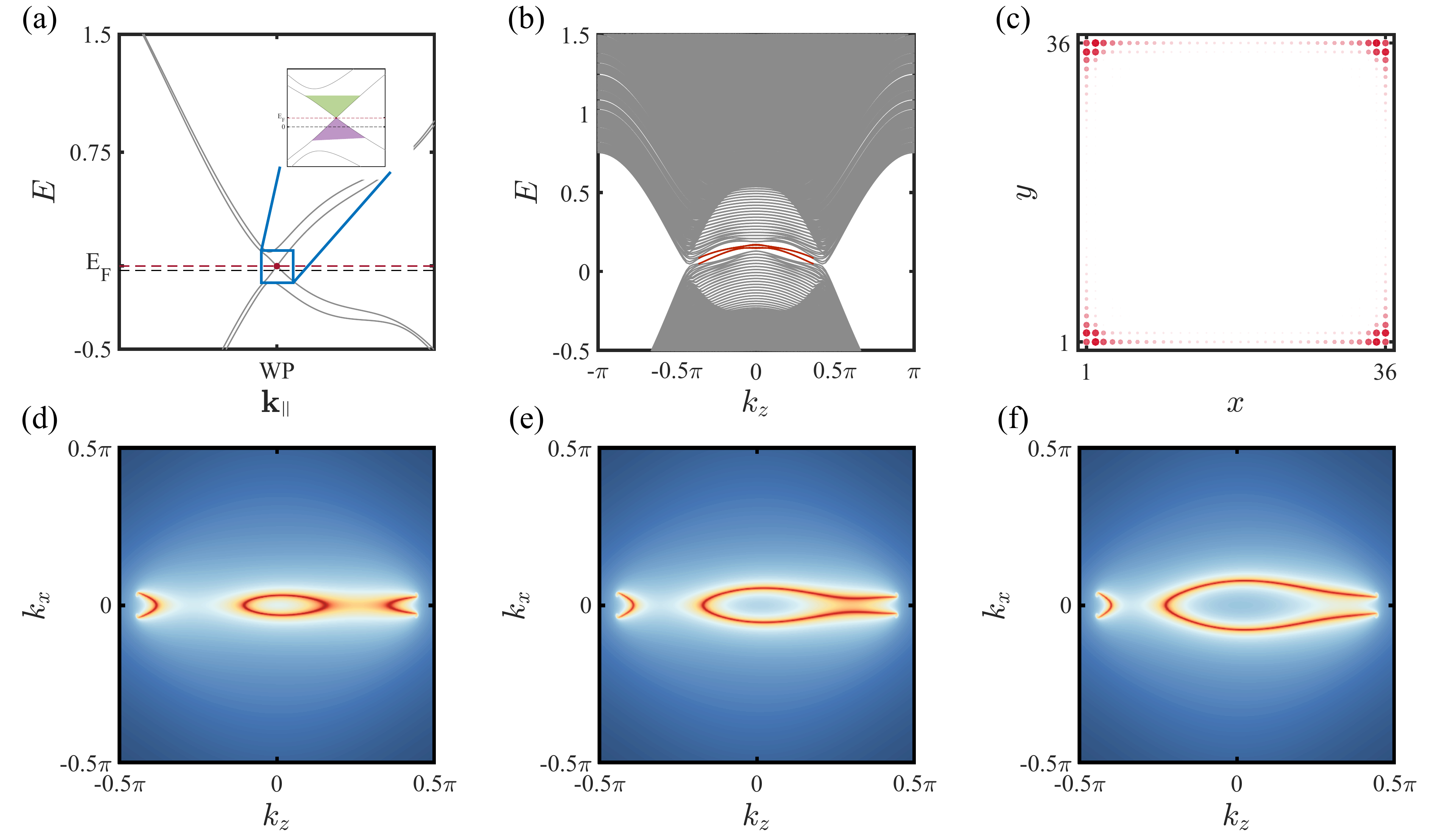}
			\caption{Floquet WSM induced by CPL propagating along the $x$-direction. (a) Bulk band around one of the Weyl nodes on the $k_x$-$k_z$ plane, marked by the red dot. Again, the green and purple triangles represent the conduction band and valence band, respectively. (b) Hinge spectrum for the open boundary condition along the $x$ and $y$ directions. The red solid lines are the gapless hinge Fermi arc states. (c) The local density of states of the hinge Fermi arc states in (b) at $k_z=0.15\pi$. $t_1=0.1$ for (a)-(c). (e-f) Evolution of surface states as $t_1$ changes. $t_1=0$ in (d), $t_1=0.05$ in (e) and $t_1=0.1$ in (f). The common parameters are chosen to be $t_2=-0.1$, $\lambda=0.5$, $t=1$, $t_z=0.8$, $K_{z}^{0}=\pi/2$, $g=-0.2$, $A=0.7$, and $\omega=3$.}
			\label{SMFigure1}
		\end{figure}

		Now, the vector potential is $\mathbf{A}(\tau)=(0,\eta\sin\omega\tau,\cos\omega\tau)$. Again, we can obtain an effective Floquet Hamiltonian which is 
		\begin{align}
			H_\text{eff}^x(\bm{k}) \nonumber & =[t_1(\mathcal{J}_{0}(A) \cos{k_z}-\cos{K_z^0})+t_2(\cos{k_x}+\mathcal{J}_{0}(A) \cos{k_y}-2)] + \lambda\sin{k_x}\Gamma_{1} \\ \nonumber
			& +\lambda\mathcal{J}_{0}(A)\sin{k_y}\Gamma_{2} \!+\![t_z(\mathcal{J}_{0}(A)\cos{k_z} \!-\!\cos{K_z^0})\!+\! t(\cos{k_x} \!+\!\mathcal{J}_{0}(A)\cos{k_y}\!-\!2)]\Gamma_{3}\\ \nonumber
			&+g(\mathcal{J}_{0}(A)\cos{k_x}\sin{k_z}-\mathcal{J}_{0}(\sqrt{2}A)\cos{k_y}\sin{k_z})\Gamma_{4} \\ \nonumber
			& +\sum\limits_{l>0,l\in odd}\frac{4\eta\mathcal{J}_{l}(A)}{l\hbar\omega}\{ -\mathcal{J}_{l}(A)\lambda t_z \cos{k_y}\sin{k_z}\Gamma_{23} +  \mathcal{J}_{l}(A) \lambda g \cos{k_x}\cos{k_y}\cos{k_z}\Gamma_{24} \\ \nonumber
			&  - \frac{\sqrt{2}}{2} \mathcal{J}_{l}(\sqrt{2}A) \lambda g \cos^{2}{k_y}\cos{k_z} \Gamma_{24} - \mathcal{J}_{l}(A) t g \cos{k_x}\sin{k_y}\cos{k_z}\Gamma_{34}  \\ 
			& + \frac{\sqrt{2}}{2} \mathcal{J}_{l}(\sqrt{2}A)[ t_z g \sin{k_y}\sin^{2}{k_z} + t g \sin{k_y}\cos{k_y}\cos{k_z} ]\Gamma_{34} \}.
			\label{S1}
		\end{align}
		Compared with the case of $z$-direction, CPL propagating along the $x$-direction not only renormalizes the original hoppings and induces the next-nearest neighbor hoppings but also generates the third-nearest even fifth-nearest neighbor remote hoppings on the $y$-$z$ plane. The induced Weyl nodes are located on the $k_x$-$k_z$ plane. The Floquet WSM has both surface states and hinge Fermi arc states as well. As discussed before, the tilt of Dirac cones in undriven Dirac semimetal~(DSM) is determined by the ratio $| t_1 / t_z |$. In the case of $z$-direction CPL, the tilt of light-induced Weyl cones is the same as that of Dirac cones as CPL does not put a correction to $| t_1 / t_z |$. In the present case, however, the tilt of Weyl cones can be tuned by CPL. When the DSM is far from the transition point, say $| t_1 / t_z |=0.125$, the induced Weyl cones are slightly tilted as shown in Fig.~\ref{SMFigure1}(a), thus the Floquet WSM belongs to type-I. Figures~\ref{SMFigure1}(b)-\ref{SMFigure1}(c) depict the hinge Fermi arc states of the WSM. Meanwhile, it supports both surface Fermi arcs and closed Fermi ring displayed in Fig.~\ref{SMFigure1}(d). Note that, the increase in $| t_1 / t_z |$ can influence the surface states in the Floquet WSM dramatically. Figures~\ref{SMFigure1}(d)-\ref{SMFigure1}(f) show the evolution of surface states as $t_1$ increases. We can see that the right surface Fermi arc and Fermi ring merge together to form a huge Fermi arc on the $k_x$-$k_z$ plane. 
		
\begin{figure}[ptb]
			\includegraphics[width=12cm]{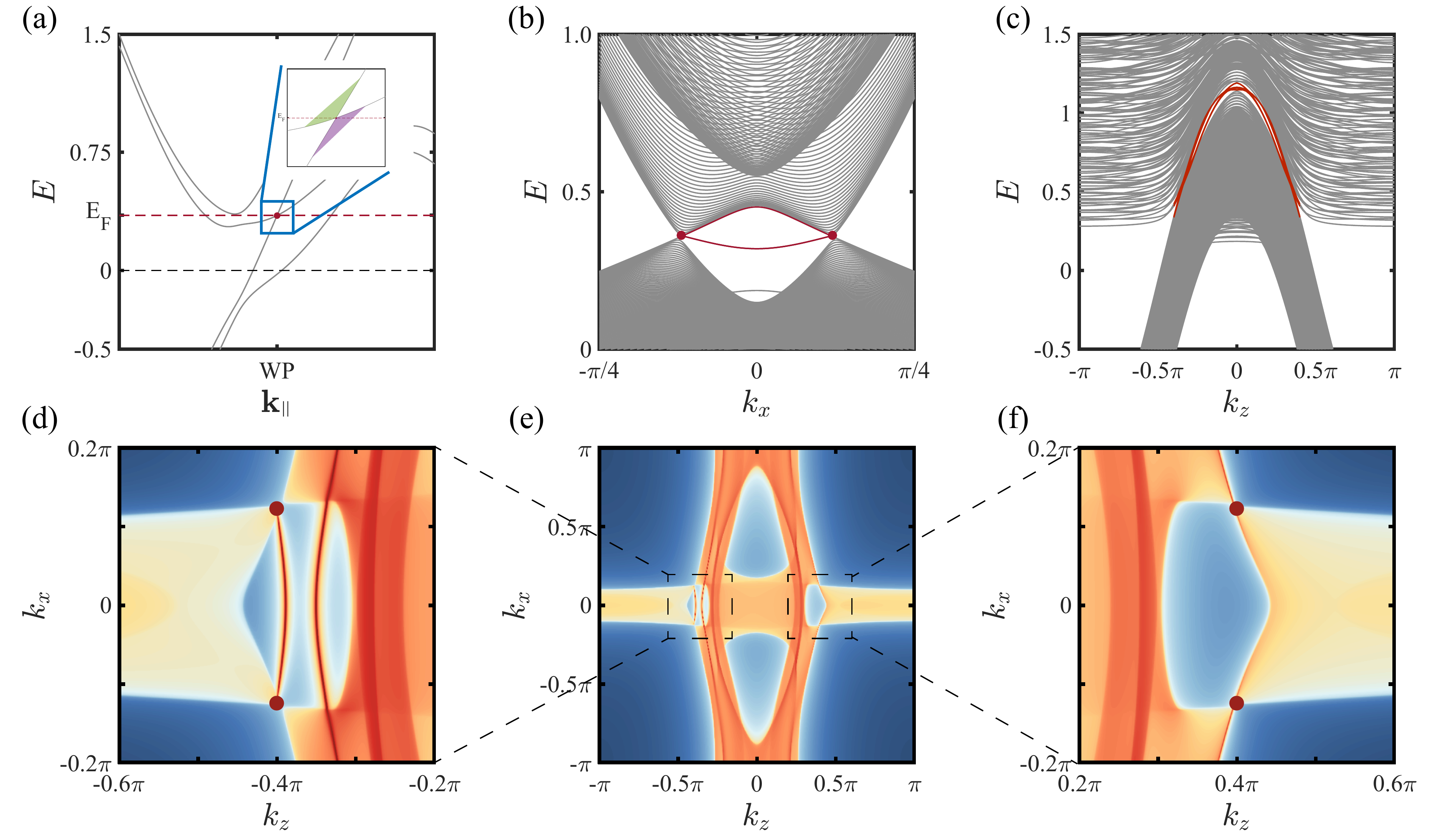}
			\caption{Energy spectrum of a type-II Floquet WSM. (a) Bulk band structure around the over-titled Weyl cones. The green and purple triangles represent the conduction band and valence band, respectively. (b) The surface band structure along $k_x$ direction. The Weyl nodes are marked by the red dots. The red solid lines mark the surface states. (c) The hinge spectrum along the $k_z$-direction. (e) The surface spectral function on the $[010]$ plane at $E=E_{F}$ marked by the red dashed line in (a). (d) and (f) are the zoom-in view of the area around two pairs of Weyl points within the black dashed squares. The parameters are the same as those in Fig.~\ref{SMFigure1} except $t_1=-t_2=0.75$, $g=-0.4$, $A=0.75$, and $\omega=1$.}
			\label{SMFigure2}
		\end{figure}

		To see how the CPL propagating along the $x$-direction generates an over-tilted type-II WSM started from a type-I DSM, we choose $| t_1 / t_z |<1$ but close to the transition point. A type-II WSM appears when tuning CPL. In Fig.~\ref{SMFigure2}(a), we plot the bulk band spectrum around one of the over-titled Weyl points. Apparently, the electron and hole pockets coexist at the Fermi level. Figures~\ref{SMFigure2}(b), \ref{SMFigure2}(d), \ref{SMFigure2}(e) and \ref{SMFigure2}(f) show the surface Fermi arc states of the type-II WSM. Notice that, there are two Fermi arcs near the left pair of Weyl nodes as shown in Fig.~\ref{SMFigure2}(d). The one, which is not terminated at the Weyl nodes, actually comes from the right pair of Weyl nodes [see Fig.~\ref{SMFigure2}(f)].  The WSM also hosts hinge Fermi arc states which are buried in the bulk states as displayed in Fig.~\ref{SMFigure2}(c). By adjusting the propagation direction of CPL, we can control the type of Floquet WSM and the surface states. Note that, when the propagation direction of CPL is along the $y$-direction, the Floquet WSM is almost the same as that induced by the CPL propagating along the $x$-direction except for the Weyl nodes are located on the $k_y$-$k_z$ plane.
		
		\section{Understanding type-II Floquet WSM transition in the continuum model}
		To understand why CPL propagating along the $x$-direction instead of the $z$-direction induces a type-II WSM  starting from a type-I DSM, we consider the low-energy continuum DSM Hamiltonian around the Dirac point locating at $\mathbf{k}^0=(0,0, K_z^0)$
		\begin{align}
			H_\text{DSM}(\bm{q}) \nonumber & =-\Big[\frac{t_1}{2}( 2 \sin{K_z^0} q_z + \cos{K_z^0} {q_z}^2 ) + \frac{t_2}{2}(q_x^2+q_y^2)\Big] + \lambda q_x \Gamma_{1} + \lambda q_y \Gamma_{2} \\
			& - \Big[ \frac{t_z}{2}( 2 \sin{K_z^0} q_z + \cos{K_z^0} {q_z}^2 ) + \frac{t}{2}(q_x^2+q_y^2) \Big]\Gamma_{3} - \frac{g}{2} (q_x^2 - q_y^2) \sin K_z^0 \Gamma_{4},
			\label{S2}
		\end{align}
		where $\bm{q}=(q_x, q_y, q_z)=\mathbf{k}-\mathbf{k}^0$ is the wave vector measured from the Dirac point. The energy dispersion at the leading (linear) order can be expressed as $E_{\pm}(\bm{q}) = T(\bm{q}) \pm U(\bm{q})$ with
		\begin{equation}
			T(\bm{q}) = - t_1 \sin{K_z^0} q_z, \qquad U(\bm{q}) =  \sqrt{{\lambda}^2 (q_x^2+q_y^2)  + \frac{1}{2} (1-\cos{2{K_z^0}}) t_z^2 q_z^2},
			\label{S3}
		\end{equation}	
		where $T(\bm{q})$ describes the tilting energy and $U(\bm{q})$ denotes the splitting of the Dirac cone. When $T$ dominates over $U$ along a specific direction, the Dirac cone is over-tilted, and the Dirac points is called type-II. In the current case, the Dirac cone is tilted along the $z$-direction. Therefore, the type of Dirac point can be determined by the ratio $|T(\bm{q}) / U(\bm{q})|=|T(q_z) / U(q_z)|=|t_1 / t_z|$. 
		
		Shinning CPL propagating along the $z$-direction, in the high-frequency regime, we obtain a correction to bare $H_\text{DSM}(\bm{q})$, which reads
		\begin{align}
			\Delta H^{z}(\bm{q}) \nonumber & =\frac{\eta A^2 e^2}{\hbar \omega} \Big[ - \frac{ \eta t \hbar \omega}{2}  \Gamma_{3} +  {\lambda}^2 \Gamma_{12} + t \lambda q_x \Gamma_{23} - t \lambda q_y \Gamma_{13} + g \lambda \sin{K_z^0} q_x \Gamma_{24} \\
			& + g \lambda \sin{K_z^0} q_y \Gamma_{14} - 2 t g \sin{K_z^0} q_x q_y \Gamma_{34} \Big].
			\label{S4}
		\end{align}
		The correction will split the Dirac point into a pair of Weyl points along the $q_z$-direction which locate at $\bm{q}^0=(0,0, Q_{z}^{1,2})$, where $Q_{z}^{1,2}=(-A^2 e^2 \hbar \omega t \pm 2 A^2 e^2 \eta {\lambda}^2)/(2 \hbar \omega t_z )$. We have fixed $K_z^0=\pi/2$ in our calculations. Meanwhile, the tilting direction is still along the $z$-direction, and the tilting ratio keeps unchanged $|T(q_z) / U(q_z)|=|t_1 / t_z|$. Therefore, CPL propagating along the $z$-direction cannot give rise to a transition to type-II WSM.
		Consider CPL propagating along the $x$-direction, the correction Hamiltonian is
		\begin{align}
			\Delta H^{x}(\bm{q}) \nonumber & =\frac{\eta A^2 e^2}{\hbar \omega} \Big{\{ }\frac{ \eta \hbar \omega}{4} \Big[ -(t_z \cos{K_z^0} + t) \Gamma_{3} + g \sin{K_z^0} \Gamma_{4} \Big] - t_z \lambda (q_z \cos{K_z^0} + \sin{K_z^0}) \Gamma_{23}  \\
			& +   g t_z \sin{K_z^0} (q_z \cos{K_z^0} + \sin{K_z^0}) q_y \Gamma_{34} \Big{\}}.
			\label{S5}
		\end{align}
		Now, the CPL induced Weyl points locate on the $q_x$-$q_z$ plane, which can be written as $\bm{q}^0=(\pm{\xi} Q_z,0,\pm Q_z)$ with  ${\xi}$ and $Q_z$ depending on the amplitude and frequency of CPL. The tilting energy becomes $T(\bm{q})={\xi} Q_z t_2 q_x +  t_1 q_z $, where ${\xi} Q_z=\sqrt{ A^2 e^2 /2 - 2 {\lambda}^2/{g^2} +\sqrt{-2 A^2 e^2 g^2 {\lambda}^2 +4 {\lambda}^4 }/{g^2}}$. After a rotation to the tilting direction, we obtain the tilting ratio
		\begin{align}
			\Big|\frac{T(q_z^{\prime})}{U(q_z^{\prime})}\Big| = \Big|\frac{t_1}{t_z}\Big| \frac{1 + ({{\xi}} {Q_z} {t_1^{\Delta}})^2 }{ \sqrt{ (1+t^{\Delta} {t_1^{\Delta}} {{\xi}}^2 {Q_z}^2 )^2 +F_A^2 } }  ,
			\label{S6}
		\end{align}
		where $t_1^{\Delta}=t_1/t_2$, $t^{\Delta}=t/t_z$, and the function $F_A $ depends on $A^2$, ${\xi}$, $Q_z$, and model parameters $t$, $\lambda$, $t_z$, and $g$. In contrast to CPL propagating along the $z$-direction, the tilting ratio can be tuned by CPL propagating along the $x$-direction. We found for the parameters used in Fig.~\ref{SMFigure2}, even $|t_1 / t_z|<1$, the additional coefficient induced by CPL will result in $|T(q_z^{\prime}) / U(q_z^{\prime})|>1$ for Weyl cones, which signals a transition to type-II WSMs. Conversely, starting from a type-II Dirac semimetal ($|t_1 / t_z|>1$), CPL can convert it into a type-I WSM.


		\bibliographystyle{apsrev4-2}

\end{document}